\documentclass{INTERSPEECH2023}

\interspeechcameraready 

\title{Integrating Emotion Recognition with Speech Recognition and \\ Speaker Diarisation for Conversations}

\name{Wen Wu$^1$\thanks{Wen Wu is supported by Cambridge International Scholarship from the Cambridge Trust. This work has been performed using resources provided by the Cambridge Tier-2 system operated by the University of Cambridge Research Computing Service funded by EPSRC Tier-2 capital grant EP/T022159/1.},Chao Zhang$^2$, Philip C. Woodland$^1$}
\address{
  $^1$Department of Engineering, University of Cambridge, Cambridge, UK\\
  $^2$Department of Electrical Engineering, Tsinghua University, Beijing, China}
\email{\{ww368, pcw\}@eng.cam.ac.uk; cz277@tsinghua.edu.cn}

\newcommand{\Loss}{\mathcal{L}}

\begin{document}

\maketitle
 
\begin{abstract}
Although automatic emotion recognition (AER) has recently drawn significant research interest, most current AER studies use manually segmented utterances, which are usually unavailable for dialogue systems. 
This paper proposes integrating AER with automatic speech recognition (ASR) and speaker diarisation (SD) in a jointly-trained system. 
Distinct output layers are built for four sub-tasks including AER, ASR, voice activity detection and speaker classification based on a shared encoder. 
Taking the audio of a conversation as input, the integrated system finds all speech segments and transcribes the corresponding emotion classes, word sequences, and speaker identities. 
Two metrics are proposed to evaluate AER performance with automatic segmentation based on time-weighted emotion and speaker classification errors.
Results on the IEMOCAP dataset show that the proposed system consistently outperforms two baselines with separately trained single-task systems on AER, ASR and SD\footnote{Code available: https://github.com/W-Wu/sTEER}.

\end{abstract}

\section{Introduction}
There has been much research work in automatic emotion recognition (AER)~\cite{ Kim2013,Poria2018,liu20b_interspeech,wu2021emotion,wu2022distribution}. However, most AER systems operate on manually segmented utterances although manual segmentations are not generally available in practical use cases. Besides, automatic speech recognition systems trained on standard speech can give poor recognition performance on emotional speech~\cite{wu2021emotion,sahu2019multi,fernandez2004computational}.

\begin{figure*}[t]
    \centering
    \includegraphics[width=0.9\linewidth]{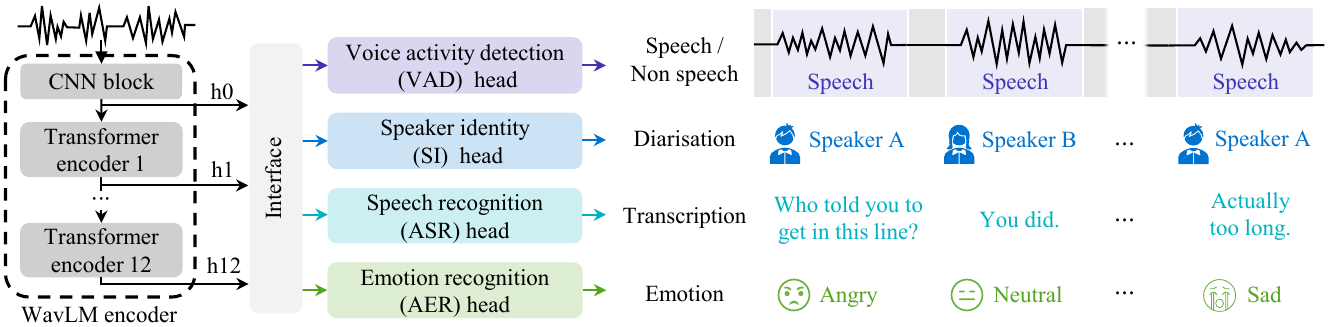}
    \vskip -1ex
    \caption{Overview of the proposed integrated system. Taking a dialogue as input, the VAD and SI head perform automatic segmentation.
    The ASR and AER head recognise the text and emotion based on the predicted segments. }
    \vspace{-2ex}
    \label{fig: overview}
\end{figure*}

This work proposes an integrated system for emotion recognition, speaker diarisation and speech recognition. Speaker diarisation is the process that detects speech regions of an audio recording and groups them into homogeneous segments according to the relative identity of the speaker. The outcome of speaker diarisation is referred to as segmentation in this paper. As illustrated in Figure~\ref{fig: overview}, the system takes the audio recording of a dialogue as input. It automatically diarises the dialogue into segments associated with different speakers, transcribes the audio segments into text, and predicts the speaker's emotion state.

The model contains four downstream heads for voice-activity-detection (VAD), speaker indentity (SI) extraction, automatic speech recognition (ASR), automatic emotion recognition and an encoder shared by all downstream heads.  
Speaker diarisation is achieved by the VAD head and the SI head. The VAD head classifies each frame into speech or non-speech. The SI head learns speaker embeddings that capture the characteristics of each speaker. 
Based on the predicted segmentation, the ASR head converts each speech segment into text and the AER head predicts the emotion states of the corresponding speaker. 
A speech foundation model~\cite{bommasani2021opportunities} (i.e. WavLM~\cite{chen2022wavlm}) is used as the shared encoder which takes raw speech waveform as input.
Foundation models are pre-trained on large amount of unlabelled data and can help handle the data sparsity issue in AER tasks.
The downstream heads take the weighted sum of intermediate hidden states from the shared encoder as input. Each head has an individual set of weights, which are trained jointly with the shared encoder and the downstream heads. 

Furthermore, this work also proposes two metrics to evaluate the emotion recognition performance with automatic segmentation: the time-weighted emotion error rate (TEER) and the speaker-attributed time-weighted emotion error rate (sTEER). To the best of our knowledge, this is the first work that considers emotion recognition with automatic segmentation and integrates emotion recognition, speech recognition and speaker diarisation into a jointly-trained model. 

The rest of the paper is organised as follows. Section~\ref{sec: related work} summarises related work. Section~\ref{sec: method} introduces the proposed system and the TEER and sTEER metrics. Section~\ref{sec: exp setup} presents the experimental setup while the results and analysis are shown in Sections~\ref{sec: results} and \ref{sec: analysis} respectively, followed by the conclusions.

\section{Related work}
\label{sec: related work}

Multi-task training and transfer learning have been investigated to improve the AER performance with ASR transcriptions. Feng  et al.~\cite{feng2020end} used an attention-based encoder-decoder model for ASR and combined the states of the ASR decoder with the acoustic features for AER. Li et al.~\cite{li2022fusing} encode the raw waveform with the Wav2Vec 2.0 model~\cite{baevski2020wav2vec} for ASR and separately as MFCCs for AER. The ASR outputs were fused into the pipeline for joint training with AER. Zhou et al.~\cite{zhou2020transfer} fine-tuned
an ASR model on emotional speech for emotion recognition. Cai et al.~\cite{cai21b_interspeech} proposed a multi-task learning framework that fed the output of a Wav2Vec 2.0 model to an ASR head and an AER head each containing one fully-connected layer. Two heads were simultaneously trained while only the AER head was kept during inference. 
Heusser et al.~\cite{heusser2019bimodal} trained ASR from audio, AER from audio and from text independently and fine-tuned the combined sub-models. Ghriss et al.~\cite{ghriss2022sentiment}
pre-trained the AER model by ASR which is trained jointly with a sentiment classifier. 

Apart from combining AER and ASR, Velichko et al.~\cite{velichko2022complex} proposed a hierarchical framework to predict gender, emotion, and deception in a cascaded way. So far, combining speaker diarisation with emotion recognition and evaluating emotion recognition when using automatic segmentation hasn't been widely studied.

\section{Proposed approach}
\label{sec: method}
\begin{figure}[tb]
    \centering
    \includegraphics[width=0.85\linewidth]{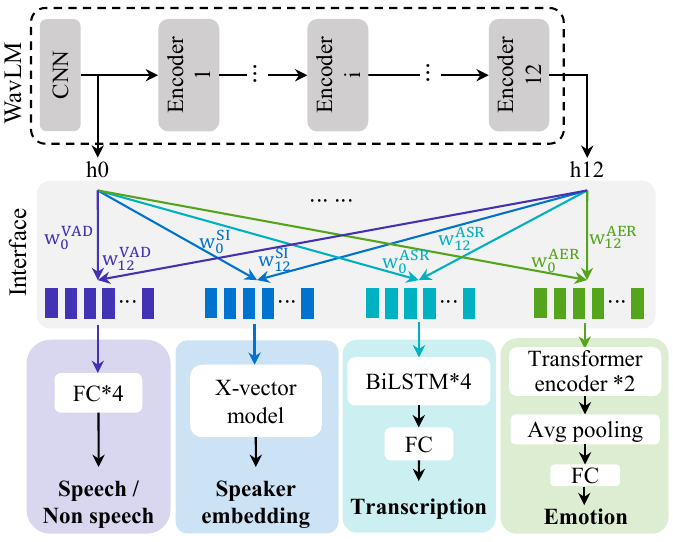}
    \vspace{-1ex}
    \caption{Structure of the proposed integrated system.}
    \label{fig: struc}
    \vspace{-2ex}
\end{figure}

The structure of the proposed system is shown in Figure~\ref{fig: struc}, which consists of a shared encoder, an interface and four downstream heads.

\subsection{Shared encoder and interface}
The WavLM Base model~\cite{chen2022wavlm} is used as the shared encoder in this paper, which takes the raw waveform as input. It contains a convolutional neural network (CNN) block as the feature extractor and 12 Transformer~\cite{vaswani2017attention} encoder blocks.
The output of the encoder is a frame sequence with a frame shift of 20~ms.

All the semantic and non-semantic information co-exist in the same speech signal. Research~\cite{chen2022wavlm,pasad2021layer,superb} has shown that intermediate representations of such foundation models contain different levels of information. 
The weighted sum of embeddings from the CNN block and each Transformer encoder block is used as the input to the downstream tasks to exploit this property. Each downstream head has its own set of weights which are trainable.

\subsection{Downstream heads}

The VAD head consists of 3 fully-connected (FC) layers with a hidden dimension of 256 and leaky ReLU activation and an output FC layer with Softmax activation which performs frame-level speech/non-speech detection. 
The SI head consists of an X-vector speaker embedding model~\cite{snyder2018x}, which generates one speaker embedding for each input sequence. The speaker embedding is fed into an FC layer with Softmax activation for speaker classification during training. During testing, spectral clustering is conducted based on speaker embeddings to produce speaker diarisation. 
The ASR head consists of 4 Bi-LSTM layers~\cite{biLSTM} with dimension of 256, 
followed by an FC layer for token prediction. A vocabulary of 29 graphemes is used including 26 letters in English plus a few punctuation characters. 
The AER head consists of 2 Transformer encoder layers of dimension 256. The representations are average-pooled along the time axis before being fed into an FC layer with Softmax activation for emotion classification. Six emotion classes are used: ``happy'', ``sad'', ``angry'', ``neutral'', ``other'', ``no majority agreement (NMA)''. ``NMA'' denotes that the human annotators don't have a majority agreed emotion class label for this utterance.

\subsection{Multi-task training loss}
The ASR head is trained using the CTC loss~\cite{CTC} and the other three heads were trained using the cross entropy loss. The shared encoder and four downstream heads are jointly trained using a multi-task loss:
\vspace{-0.5ex}
\begin{equation}
    \Loss_{\text{Total}} = \epsilon_{\text{VAD}}\Loss_{\text{VAD}}+\epsilon_{\text{SI}}\Loss_{\text{SI}}+\epsilon_{\text{ASR}}\Loss_{\text{ASR}}+\epsilon_{\text{AER}}\Loss_{\text{AER}}
    \label{eq: MTL loss}
\end{equation}
where $\epsilon_{\text{VAD}},\epsilon_{\text{SV}}, \epsilon_{\text{ASR}},\epsilon_{\text{AER}}$ are coefficients that are set manually to keep the weighted loss of the three head in the same scale.

\subsection{Training and testing procedures}
Teacher forcing is applied during training using reference segmentations. The system takes segmented utterances as input. The encoder and downstream heads are trained jointly using the multi-task loss defined in Eqn~\ref{eq: MTL loss}. The segmented utterances can contain silence at the beginning, between words and at the end. The VAD head is trained based on the intra-utterance silence.

During testing, dialogues are input into the system. The system can benefit from knowing the context. A sliding window of \SI{3}{\second} length and \SI{1}{\second} overlap is applied to the dialogue. VAD is performed on each window.
To avoid overlapped regions being counted twice at the output, the results of the middle second was kept for each window. This is equivalent to taking previous \SI{1}{\second} and future \SI{1}{\second} as context when making a prediction on the current \SI{1}{\second} of audio data.  Post-processing is applied to the VAD predictions, so speech/non-speech regions shorter than \SI{0.25}{\second} are removed. 
A smaller sliding window of \SI{1}{\second} length and \SI{0.5}{\second} overlap is then applied to the detected speech regions.
The SI head extracts a single speaker embedding from each window. Spectral clustering is used based on the speaker embeddings which groups segments from the same speaker together, thus producing an automatic segmentation. 
Based on the automatic segmentation, the ASR and AER heads takes each segment as input and predict the text and emotion respectively.

\subsection{Evaluating emotion classification performance with automatic segmentation}
\label{sec: sTEER}
The segments predicted by the system can have different start and end times to the reference segments.  Therefore, the classification accuracy is no longer sufficient to evaluate the performance of emotion classification in this case since it cannot handle the alignment between segments. This paper, therefore, proposes the time-weighted emotion error rate (TEER) in order to evaluate the AER performance given non-oracle segmentations. The TEER is computed as follows:
\begin{equation}
    \text{TEER} = \frac{\text{MS}+\text{FA}+\text{CONF}_\text{emo}}{\text{TOTAL}}
    \label{eq: TEER}
\end{equation}
where missed speech (MS) is the duration of speech incorrectly classified as non-speech, false alarm speech (FA) is the duration of non-speech incorrectly classified as speech, confusion (CONF$_\text{emo}$) is the audio duration where emotion is wrongly classified, and TOTAL is the sum of the reference speech duration for all utterances.

Furthermore, the speaker-attributed TEER (sTEER) is proposed which expects the system to accurately predict both the speaker and their emotion. The sTEER is computed as follows:
\begin{equation}
    \text{sTEER} = \frac{\text{MS}+\text{FA}+\text{CONF}_\text{emo+spk}}{\text{TOTAL}}
    \label{eq: sTEER}
\end{equation}
CONF$_\text{emo}$ in Eqn~\ref{eq: TEER} is replaced by CONF$_\text{emo+spk}$, which is the duration where either speaker or emotion is wrong. sTEER reflects the overall performance of both speaker diarisation and emotion classification.

\section{Experimental setup}
\label{sec: exp setup}
\subsection{Dataset}
The benchmark IEMOCAP dataset~\cite{Busso2008IEMOCAP} was used which consists of approximately 12 hours of English speech including 5 dyadic conversational sessions. 
There are in total 151 dialogues including 10,039 utterances. IEMOCAP provides the time-stamp of each utterance in a dialogue as well as word-level alignments of each utterance. The alignments show that 40\% frames in the segmented utterances are silence.
Each utterance was annotated by three human annotators. Sentences that don't have majority agreed emotion label from the annotators accounts for a 25\% of the dataset.

AER in this paper uses a six-way classification setup. Emotion class ``excited'' is merged with ``happy''. All sentences with emotion label other than `` happy'', ``sad'', ``angry'', ``neutral'' are grouped into class ``others''.
Sentences that don't have a majority agreed emotion label from the annotators are grouped into the sixth class ``NMA''.
Speaker exclusive leave-one-session-out five-fold cross validation (CV) are performed and the average results are reported.
Speakers in the test set are unseen in the training and validation set and it is ensured that utterances from the same dialogue are either all in the training set or all in the validation set.

\subsection{Evaluation metrics}
The false alarm rate (FAR) and missed rate (MSR) were used to evaluate the VAD performance. FAR computes the ratio of the number of non-speech frames mispredicted as speech to the total number of speech frames. MSR computes the ratio of the number of speech frames mispredicted as non-speech to the total number of speech frames. Diarisation error rate (DER) was used to evaluate the performance of diarisation which maps the predicted relative speaker identity to the true speaker identity and measures the fraction of time not attributed correctly to a speaker or to non-speech. Overlapped speech was considered when computing DER. Since manual annotations cannot be precise at the audio sample level, it is common to remove from evaluation a forgiveness collar around each segment boundary. Unless otherwise mentioned, a collar of \SI{0.25}{\second} was applied when evaluating with automatic segmentation.

Word error rate (WER) and classification accuracy (Acc$_\text{emo}$) were used to evaluate the performance of speech recognition and emotion classification respectively with oracle segmentation. With automatic segmentation, the concatenated minimum-permutation word error rate (cpWER)~\cite{watanabe2020chime} was used to evaluate the ASR system performance which concatenates utterances of the same speaker and computes the WER.  
The sTEER and TEER were used to evaluate the AER system which have been defined in Section~\ref{sec: sTEER}.

\subsection{Training specification}
The shared encoder was initialised with the  publicly available WavLM Base+ model\footnote{{https://huggingface.co/microsoft/wavlm-base-plus}}. It was fine-tuned jointly with the downstream head while the CNN feature extractor of the WavLM model was frozen during fine-tuning. 
Speed perturbation was applied to the ASR and SI heads. For each epoch, the speed of each waveform was randomly adjusted to 0.95 or 1.05 of the original speech or remain unchanged. Speed perturbation was not applied to AER since speed is an important clue for emotion detection. The model was implemented using the SpeechBrain toolkit~\cite{ravanelli2021speechbrain}.
The system was trained using Adadelta optimiser with the Newbob learning rate scheduler.
Scaling coefficients $\epsilon_{\text{VAD}}$ and $\epsilon_{\text{SI}}$ were set to 1.2 while the other two were set to 1.   
For each fold in 5-CV, training took around 5 hours on an NVIDIA A100 GPU and five checkpoints with the lowest validation loss were averaged after training for testing.

\section{Results}
\label{sec: results}
The performance of the SI, ASR and AER heads were first evaluated with oracle segmentations in Section~\ref{sec: oracle results}. The complete system was then evaluated with automatic segmentation in Section~\ref{sec: auto results}.  
The proposed system was compared to two baseline models:
\begin{itemize}
    \item ``Baseline-ref": Reference models which have been trained on other large datasets. The reference ASR model was pre-trained using 100 hours of LibriSpeech~\cite{panayotov2015librispeech} training data, which has a WER of 5.64\% on ``test-clean'' set  and 12.15\% on ``test-other'' set. The reference speaker embedding model\footnote{{https://huggingface.co/microsoft/wavlm-base-plus-sv}} was pre-trained on Voxceleb 1.0~\cite{VoxCeleb}. A VAD module~\cite{sun2021combination} pre-trained on the augmented multi-party interaction (AMI) meeting corpus~\cite{carletta2006ami} was used as the reference baseline for VAD, which has 2.1\% FAR and 4.7\% MSR on the AMI eval set. No reference model was used for AER since we are evaluating on the emotion dataset. The reference system is a cascade of the reference models in the order of VAD, SI, ASR.
    \item ``Baseline-frzn": The shared encoder was frozen during training. In this case, the four downstream heads were independent of each other and the multi-task loss becomes equivalent to training each head separately. 
\end{itemize}

\begin{table}[t]
\centering
\caption{Results with reference segmentation. Collar was set to 0 when computing DER since oracle segmentation was assumed. `$\uparrow$' denotes the higher the better, `$\downarrow$' denotes the lower the better. Best results of each column are shown in bold.}
\vspace{-1ex}
\begin{tabular}{c|ccc}
\toprule
            & \textbf{\%Acc$_\text{emo}$$\uparrow$} & \textbf{\%WER$\downarrow$}     &\textbf{\%DER$\downarrow$ }\\
\midrule
Baseline-ref  & / & 33.29 & 1.10        \\
Baseline-frzn & 44.44 & 31.43 & 0.40 \\
\midrule
Proposed        & \textbf{49.49}  & \textbf{24.61} & \textbf{0.30} \\
\bottomrule
\end{tabular}
\label{tab: oracle}
\end{table}

\begin{table}[t]
\centering
\caption{VAD and speaker diarisation results.}
\vspace{-1.5ex}
\begin{tabular}{c|cc|c}
\toprule
& \textbf{\%FAR$\downarrow$} & \textbf{\%MSR$\downarrow$} & \textbf{\%DER$\downarrow$}    \\
\midrule
Baseline-ref  & 5.15&        1.30&  8.20\\
Baseline-frzn    & 3.14&        1.16&  7.04\\
\midrule
Proposed         & \textbf{2.91}&        \textbf{1.06}&  \textbf{6.87}\\
\bottomrule
\end{tabular}
\label{tab: VAD}
\vspace{-1.5ex}
\end{table}

\begin{table}[b]
\centering
\vspace{-1ex}
\caption{Speaker-attributed ASR and AER performance under automatic segmentation. Best results shown in bold.}
\label{tab: sTEER}
\vspace{-1.5ex}
\begin{tabular}{c|c|cc}
\toprule
 &  \textbf{\%cpWER$\downarrow$} & \textbf{\%sTEER$\downarrow$} & \textbf{\%TEER$\downarrow$} \\
\midrule
Baseline-ref& 43.76     & / & / \\
Baseline-frzn& 41.19 &	69.49    & 68.70  \\
\midrule
Proposed   & \textbf{36.20} &	\textbf{66.03}& \textbf{65.17}\\
\bottomrule
\end{tabular}

\end{table}

\subsection{Oracle segmentation}
\label{sec: oracle results}

In this section, utterances based on the reference segmentation were used as input to the SI, ASR and AER heads.
The results are shown in Table~\ref{tab: oracle}. The ASR head with a frozen encoder reduced the WER on IEMOCAP to 31.43\%
and the SI head with frozen encoder reduced the DER to 0.40\%.
The proposed system with shared encoder jointly fine-tuned with downstream heads further reduced the WER and DER to 24.61\% and 0.30\% respectively. The 6-way emotion classification accuracy increased from 44.44\% to 49.49\%.
The proposed integrated system outperforms the baselines for all three heads. Fine-tuning the pre-trained encoder on emotion data helps to adapt it to the specific domain, while sharing the encoder between the four downstream heads helps to capture general information relevant to the domain and avoids overfitting to trivial patterns, especially given the scarcity of data.

\subsection{Automatic segmentation}
\label{sec: auto results}
The VAD performance and diarisation results based on VAD predictions are summarised in Table~\ref{tab: VAD}. The proposed system produced the best results on both VAD and diarisation.

ASR and AER were conducted using the diarisation output. As shown in Table~\ref{tab: sTEER}, the proposed integrated system reduced cpWER by relative 12\% compared to the baselines. sTEER is slightly higher than TEER as it takes speaker prediction error into account. The proposed system outperforms the single AER head in both emotion metrics, showing its superior performance for  emotion recognition with automatic segmentation.

\section{Discussion and analysis}
\label{sec: analysis}
\subsection{Trainable weights of the interface}
\begin{figure}[t]
    \centering
    \includegraphics[width=\linewidth]{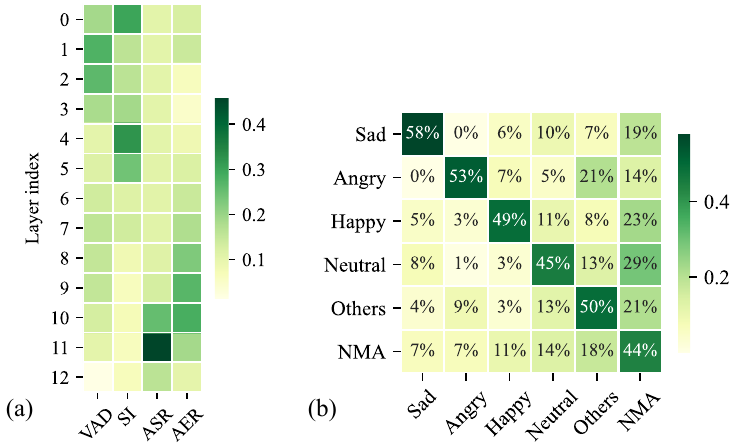} 
    \vspace{-3ex}
    \caption{(a) Weights of the interface for different downstream heads. (b) Confusion matrix of 6-way emotion recognition.}
    \label{fig: heatmap}
    \vspace{-3ex}
\end{figure}

The trainable weights for the four downstream heads are plotted in Figure~\ref{fig: heatmap}(a). As can be seen, layer 0 and layer 4 are particularly useful for extracting speaker information. Layer 8-10 are more effective for AER and layer 11 contains most text information. This shows a similar pattern to previous findings~\cite{chen2022wavlm,pasad2021layer} that block-wise evolution of intermediate representations of a foundation model follows an acoustic-linguistic hierarchy, where the lower layers encode speaker-related information and higher layers encode phonetic/semantic information.

\subsection{Confusion matrix of 6-way emotion classification}
\label{sec: confusion matrix}
Based on the 6-way predictions, 
4-way classification accuracy considering ``happy'', ``sad'', ``angry'', ``neutral'' is 73.96\%, which is better than the results of Wav2Vec 2.0 Base (63.43\%) and WavLM Base+ (68.65\%) from the SUPERB leaderboards~\cite{superb}. The class ``NMA'' is relatively easily confused, as shown by the 6-way confusion matrix of in Figure~\ref{fig: heatmap}(b). For utterances classified as ``NMA'', the human annotators gave different emotion class labels and didn't reach majority agreement. These utterances may contain ambiguous emotions, mixed emotions, or emotions that tend to confuse the annotators. Among the other five classes, ``angry'' is the least likely to be confused with ``NMA'' probably because ``angry'' is relatively less ambiguous. By contrast, ``neutral'' is more likely to be wrongly predicted as ``NMA'', possibly because neutral emotions are relatively weak and human annotators are likely to disagree due to subjective perception.

\section{Conclusions}
This paper proposes a system that integrates  emotion recognition with speech recognition and speaker diarisation in a jointly-trained model.
The system investigates emotion recognition with automatic segmentation to address the issue of lacking manual segmentation in practical applications. 
The system also improves recognition performance on emotional speech by 12\% reduction in relative  word error rate with automatic segmentation. 
Time-weighted emotion error rate and speaker-attributed time-weighted emotion error rate were proposed to evaluate emotion classification performance when segmentation is non-oracle. Although the benchmark dataset used in this paper contains only dyadic conversations, the proposed method can also be applied to multi-party conversations.

\bibliographystyle{IEEEtran}
\bibliography{mybib}

\end{document}